\documentclass[12pt]{article}

\begin{document}

\title{Dynamical Correlations \\ for Vicious Random Walk \\ with a Wall}

\author{Taro Nagao}
\date{}
\maketitle

\begin{center}
\it
Institut f\"ur Theoretische Physik, Universit\"at zu K\"oln, \\ 
Z\"ulpicher Str. 77, 50937 K\"oln, Germany \\  
(Permanent address: Department of Physics, Graduate 
School of Science, \\ Osaka University, Toyonaka, Osaka 
560-0043, Japan) 
\end{center}

\bigskip
\begin{center}
\bf Abstract 
\end{center}
\par
\bigskip
\noindent
A one-dimensional system of nonintersecting Brownian particles 
is constructed as the diffusion scaling limit of Fisher's 
vicious random walk model. $N$ Brownian particles start from the 
origin at time $t=0$ and undergo mutually avoiding motion 
until a finite time $t=T$. Dynamical correlation 
functions among the walkers are exactly evaluated 
in the case with a wall at the origin. 
Taking an asymptotic limit $N \rightarrow \infty$, we observe 
discontinuous transitions in the dynamical correlations. 
It is further shown that the vicious walk model with a wall 
is equivalent to a parametric random matrix model describing 
the crossover between the Bogoliubov-deGennes 
universality classes. 
 
\newpage
\noindent
\section{Introduction}
\setcounter{equation}{0}
\renewcommand{\theequation}{1.\arabic{equation}}

The vicious walk model was first introduced by Fisher 
and applied to wetting and melting phenomena\cite{MEF,PJF}. 
It recently attracts renewed interest in statistical 
and mathematical physics,  
since intimate connections were established to other research fields, 
such as Young tableaux in combinatorics\cite{GOV,BDJ,KGV}, asymmetric 
exclusion process (ASEP) in nonequilibrium statistical mechanics\cite{KJ1}, 
Kardar-Parisi-Zhang (KPZ) universality in surface growth 
process\cite{PS1,PS2,KJ2} and the theory of random matrices\cite{BAIK,NFV}. 
In the context of random matrix theory, the ensembles of vicious 
walkers in one dimension correspond to discretizations of random 
matrix ensembles. 
\par
Suppose that there are $N$ walkers on a lattice 
${\bf Z} = \{\cdots,-2,-1,0,1,2,\cdots \}$. Starting 
from $N$ distinct (even integer) positions 
$2 s_1 < 2 s_2 < \cdots < 2 s_N$, 
at each time step each walker moves to the left or right one 
lattice site with equal probability. Let us denote the position 
of the $j$-th walker at time $k \geq 0$ by $R^{s_j}_k$. Walkers are 
"vicious" so that two or more walkers are prohibited to arrive at 
the same site simultaneously. That is, the nonintersecting condition 
\begin{equation}
R^{s_1}_k < R^{s_2}_k < \cdots < R^{s_N}_k, \ \ \ 1 \leq \forall k \leq K
\end{equation}
is imposed. In this paper, we further impose a condition
\begin{equation}
R^{s_j}_k \geq 0, \ \ \ 1 \leq \forall k \leq K,
\end{equation}
which implies that there is a wall at the origin. Let us define 
$V(R^{s_j}_K=2 d_j)$ as the realization probability that the vicious 
walkers arrive at (even integer) 
positions $2 d_1 < 2 d_2 < \cdots < 2 d_N$ at 
an even integer time $K$. Utilizing the Lindstr\"om-Gessel-Viennot 
theorem and the reflection principle of random walks, we find an 
explicit formula\cite{KGV}
\begin{equation}
V(R^{s_j}_K=2 d_j) = 2^{-KN} \det\left[ \left( \frac{K}{K/2 + s_k - d_j} 
\right) - \left( \frac{K}{K/2 + s_k + d_j  + 1} \right) \right]_{j,k=1,2,\cdots,N}.
\end{equation}
This realization probability can be simplified in the diffusion 
scaling limit\cite{KT1,KT2,NKT,KTM}. Let us introduce a 
positive number $L$ and set $K=Lt$, 
$s_j=\sqrt{L} x_j/2$ and $d_j = \sqrt{L} y_j/2$. Then we can easily find
\begin{eqnarray}
& & f(t;y_1,y_2,\cdots,y_N \mid x_1 x_2,\cdots,x_N) 
\equiv \lim_{L \rightarrow \infty} 
\left(\frac{\sqrt{L}}{2}\right)^N V(R^{\sqrt{L}x_j/2}_{Lt} = 
\sqrt{L} y_j) \nonumber \\ 
& = & (2 \pi t)^{-N/2} \det\left[
{\rm exp}\left\{ - \frac{1}{2 t} (x_j - y_k)^2 \right\} - 
{\rm exp}\left\{ - \frac{1}{2 t} (x_j + y_k)^2 \right\} 
\right].
\end{eqnarray}
This function gives the nonintersecting probability of the Brownian 
particles on the rescaled lattice ${\bf Z}/(\sqrt{L}/2)$ up to times $t$ 
depending on the initial positions $\{x_j\}$ and final positions $\{y_j\}$. 
Therefore the probability amplitude that the vicious walkers are located at 
$x^j_1,x^j_2,\cdots,x^j_N$ at times $t_j$, 
$j=0,1,2,\cdots,M$, is given by 
\begin{eqnarray}
& & P(t_0;x^0_1,x^0_2,\cdots,x^0_N;t_1;x^1_1,x^1_2,\cdots,x^1_N;
\cdots;t_M;x^M_1,x^M_2,\cdots,x^M_N) 
\nonumber \\ & = & 
\prod_{j=0}^{M-1} \varphi^{T}(t_j;x^j_1,x^j_2,\cdots,x^j_N;
t_{j+1};x^{j+1}_1,x^{j+1}_2,\cdots,x^{j+1}_N),
\end{eqnarray}
where 
\begin{eqnarray}
& & \varphi^T(s;x_1,x_2,\cdots,x_N;t;y_1,y_2,\cdots,y_N) 
\nonumber \\ & = & f(t-s;y_1,y_2,\cdots,y_N \mid x_1,x_2,\cdots,x_N) 
\frac{{\cal N}(T-t;y_1,y_2,\cdots,y_N)}{
{\cal N}(T-s;x_1,x_2,\cdots,x_N)}, \nonumber \\ 
\end{eqnarray}
\begin{equation}
{\cal N}(t;x_1,x_2,\cdots,x_N) = \int_{0<y_1<y_2<
\cdots<y_N<\infty} 
{\rm d}y_1 {\rm d}y_2 \cdots {\rm d}y_N f(t;y_1,y_2,\cdots,y_N \mid 
x_1,x_2,\cdots,x_N).
\end{equation}
The dynamical correlation functions among the walkers 
at times $t_1,t_2,\cdots,t_M$ are defined as 
\begin{eqnarray} 
& & \rho(x^1_1,\cdots,x^1_{n_1};x^2_1,\cdots,x^2_{n_2};
\cdots;x^M_1,\cdots,x^M_{n_M}) 
\nonumber \\ & = & \frac{1}{\prod_{l=1}^M (N - n_l)!} 
\int_0^{\infty} \prod_{j=1}^N 
{\rm d}x^0_j \int_0^{\infty} \prod_{j=n_1 + 1}^N {\rm d}x^1_j 
\cdots \int_0^{\infty} \prod_{j=n_M + 1}^N {\rm d}x^M_j \nonumber \\ 
& \times & p_0(\{x^0_j\}) \prod_{m=0}^{M-1} \varphi^{T}(t_m,\{x^m_j\};t_{m+1}, 
\{x^{m+1}_j\}).
\end{eqnarray}
Here $p_0(\{x^0_j\})$ is the initial distribution at $t_0=0$. 
Let us suppose that all the Brownian particles start at 
the origin so that we can set $p_0(\{x^0_j\}) = \prod_j \delta(x^0_j)$. 
Making replacements
\begin{equation}
x^1_j \rightarrow \sqrt{y^1_j}, \ \ 
x^2_j \rightarrow \sqrt{y^2_j}, \cdots,
x^M_j \rightarrow \sqrt{y^M_j}, \ \ 
x^{M+1}_j \rightarrow \sqrt{y^{M+1}_j}, 
\end{equation}
we rewrite the dynamical correlation functions as
\begin{eqnarray}
& & \rho(y^1_1,\cdots,y^1_{n_1};y^2_1,\cdots,y^2_{n_2};\cdots;y^M_1,
\cdots,y^M_{n_M}) \nonumber \\ 
& \propto & \int_0^{\infty} \prod_{j=n_1+1}^N {\rm d}y^1_j
\int_0^{\infty} \prod_{j=n_2+1}^N {\rm d}y^2_j 
\cdots \int_0^{\infty} \prod_{j=n_M+1}^N {\rm d}y^M_j 
\int_0^{\infty} \prod_{j=1}^N {\rm d}y^{M+1}_j 
\nonumber \\ & \times & 
\prod_{j>k}^N (y^1_j - y^1_k) 
\prod_{j>k}^N {\rm sgn}(y^{M+1}_j - y^{M+1}_k) 
\prod_{m=1}^M {\rm det}\left[ g^m(y^m_j,y^{m+1}_k) 
\right]_{j,k=1,2,\cdots,N}. 
\label{RHOINT} 
\end{eqnarray}
Here
\begin{eqnarray} 
& & g^1(x,y) = \frac{1}{2 \sqrt{2 \pi t_1}} 
\frac{1}{\sqrt{2 \pi (t_2-t_1) y}} 
{\rm exp}\left\{- \frac{1}{2 t_1} x \right\} 
\nonumber \\ & \times &  
\left[ {\rm exp}\left\{ - \frac{1}{2 (t_2-t_1)} 
(\sqrt{x} - \sqrt{y})^2\right\} - {\rm exp}\left\{ 
- \frac{1}{2 (t_2-t_1)} (\sqrt{x} 
+ \sqrt{y})^2\right\} \right] \nonumber \\ 
\end{eqnarray} 
and
\begin{eqnarray}
& & g^m(x,y) = \frac{1}{2 \sqrt{2 \pi (t_{m+1}-t_m) y}} 
\nonumber \\ & \times &  
\left[ {\rm exp}\left\{- \frac{1}{2 (t_{m+1}-t_m)} 
(\sqrt{x} - \sqrt{y})^2 \right\} - {\rm exp}\left\{ 
- \frac{1}{2 (t_{m+1}-t_m)} (\sqrt{x} + \sqrt{y})^2 \right\} \right], 
\nonumber \\ & & 2 \leq m \leq M. \nonumber 
\end{eqnarray}
\par
In the case with no wall, the vicious walk model in the diffusion 
scaling limit is equivalent to the eigenvalue dynamics of parametric 
random matrices belonging to the standard symmetry 
class\cite{KT2,NKT}. Similarly, in the presence of a wall, it will 
be shown that the vicious walk model and parametric 
random matrices with the Bogoliubov-deGennes symmetry 
are equivalent. The Bogoliubov-deGennes matrix model 
was proposed by Altland and Zirnbauer as an effective 
model of mesoscopic normalconducting-superconducting 
hybrid structures\cite{AZ1,AZ2}. 
\par
This paper is organized as follows. In \S 2, quaternion determinant 
expressions for the dynamical correlation functions are presented. 
In \S 3, we rewrite the quaternion determinant expressions 
in terms of the Laguerre polynomials. In \S 4, asymptotic 
forms of the dynamical correlation functions are evaluated 
in the limit $N \rightarrow \infty$. In \S 5, an equivalence 
between the vicious walk and the Bogoliubov-deGennes matrix 
model is demonstrated.

\section{Dynamical Correlation Functions}
\setcounter{equation}{0}
\renewcommand{\theequation}{2.\arabic{equation}}

\subsection{Quaternion Determinant Expressions}
We begin with the definition of a quaternion determinant\cite{DYQ}. 
A quaternion is defined as a linear combination of 
four basic units $\{1, e_1, e_2, e_3 \}$:
\begin{equation} 
q=q_0+q_1e_1+q_2e_2+q_3e_3. 
\end{equation}
Here $q_0, q_1, q_2$ and $q_3$ are real or complex numbers. 
We call $q_0$ the scalar part of $q$. 
The quaternion multiplication is associative 
but in general not commutative. The multiplication rule of 
the four basic units are given by
\begin{equation} 
1 \cdot 1=1,\;\; 1 \cdot e_j=e_j \cdot 
1=e_j,\;\; j=1,2,3, \nonumber 
\end{equation}
\begin{equation} 
e_1^2=e_2^2=e_3^2=e_1e_2e_3=-1. 
\end{equation} 
We define a dual ${\hat q}$ a quaternion $q$ 
\begin{equation} 
{\hat q}=q_0-q_1e_1-q_2e_2-q_3e_3. 
\end{equation} 
A dual is an analogue of the complex conjugate of a complex 
number. For a matrix $Q$ with quaternion elements $q_{jl}$, 
we can also define a dual matrix ${\hat Q}=[{\hat q}_{lj}]$. 
The quaternion units can be represented as $2 \times 2$ matrices
\begin{displaymath} 
1 \rightarrow 
\left[ \begin{array}{cc} 1 & 0 \\ 0 & 1 \end{array} 
\right], \ \ e_1 \rightarrow \left[ \begin{array}{cc} 0 & -1 \\ 1 & 0 \end{array} \right],
\end{displaymath}
\begin{equation} 
e_2 \rightarrow \left[ \begin{array}{cc} 0 & -i \\ -i & 0 \end{array} \right], 
\ \ e_3 \rightarrow \left[ \begin{array}{cc} i & 0 \\ 0 & -i 
\end{array} \right]. 
\end{equation} 
\par
Let us now introduce a quaternion determinant {\rm Tdet}. 
For a self-dual $Q$ ( i.e., $Q={\hat Q}$ ), it is defined as 
\begin{equation} {\rm Tdet}\, Q = \sum_P (-1)^{N-l} \prod_1^l (q_{ab} 
q_{bc} \cdots q_{da})_0.
\end{equation}
Here $P$ denotes any permutation of the indices 
$(1,2,\cdots,N)$ consisting of $l$ exclusive cycles of the form 
$(a \rightarrow b \rightarrow c \rightarrow \cdots \rightarrow d 
\rightarrow a)$ and $(-1)^{N-l}$ is the parity of $P$. 
The subscript $0$ means that the scalar part of 
the product is taken over each cycle. If 
all the elements of $Q$ have only the scalar parts, 
every element is commutable so that a quaternion 
determinant becomes an ordinary determinant.
\par
In both cases $N$ even and odd, we define 
quaternion matrices $B^{\mu \nu}$, $\mu,\nu=1,2,\cdots,M$ 
in terms of the following $2 \times 2$ 
representations of the quaternion elements:
\begin{equation}
B^{\mu \nu}_{jl} =  \left[ \begin{array}{cc} {\tilde S}^{\mu \nu}_{jl} & 
{\tilde I}^{\mu \nu}_{jl} \\ 
D^{\mu \nu}_{jl} & {\tilde S}^{\nu \mu}_{lj} \end{array} \right], 
\ \ j,l = 1,2,\cdots,N.
\end{equation}
The matrices ${\tilde S}^{\mu \nu}$,${\tilde I}^{\mu \nu}$ and 
$D^{\mu \nu}_{jl}$ are defined in the following subsections.    
Applying the integration technique developed in Ref.\cite{TN} to the 
integral (\ref{RHOINT}), we find the quaternion determinant expression 
of the dynamical correlation functions
\begin{eqnarray}
& & \rho(y^1_1,\cdots,y^1_{n_1};y^2_1,\cdots,y^2_{n_2};\cdots;y^M_1,
\cdots,y^M_{n_M}) = {\rm Tdet}[B^{\mu \nu}(n_{\mu},n_{\nu})], 
\nonumber \\ & & \mu,\nu=1,2,\cdots,M,
\end{eqnarray}
where each block $B^{\mu \nu}(n_{\mu}, n_{\nu}) $ is obtained by removing the $n_{\mu} + 1, n_{\mu} + 2,  \cdots, N$-th rows and $n_{\nu} + 1, n_{\nu} + 2, \cdots, N$-th columns from $B^{\mu \nu}$.  

\subsection{Skew Orthogonal Polynomials}
In order to define the matrices ${\tilde S}^{\mu \nu}$,
${\tilde I}^{\mu \nu}$ and $D^{\mu \nu}_{jl}$, we need 
to first introduce skew orthogonal polynomials. In terms of
\begin{eqnarray}
& & G^{mn}(x,y) \nonumber \\ 
& = & \left\{ \begin{array}{l} \delta(x-y), \ \ m=n, \\ 
g^m(x,y), \ \ m=n-1, \\  
\displaystyle \int_0^{\infty} {\rm d}y_{m+1} 
{\rm d}y_{m+2} \cdots {\rm d}y_{n-1} \\
\times g^m(x,y_{m+1}) g^{m+1}(y_{m+1},y_{m+2}) 
\cdots g^{n-1}(y_{n-1},y), \ \ 
m < n - 1, \end{array} \right. \nonumber \\  
\end{eqnarray}
we define 
\begin{eqnarray}
F^{mn}(x,y) & = & \int_0^{\infty} {\rm d}z^{\prime} 
\int_0^{z^{\prime}} {\rm d}z   
\{ G^{m \ M+1}(x,z) G^{n \ M+1}(y,z^{\prime}) \nonumber \\ 
& - & G^{n \ M+1}(y,z) G^{m \ M+1}(x,z^{\prime}) \}.
\end{eqnarray}
\par
In terms of an antisymmetric inner product
\begin{equation}
 \langle f(x), g(y) \rangle_m = \frac{1}{2} \int_0^{\infty} 
{\rm d}x \int_0^{\infty} {\rm d}y  
F^{mm}(y,x)  [f(y) g(x) - f(x) g(y) ],
\end{equation}
monic polynomials $R^1_k(x)=x^k+\cdots$ of 
degree $k$ are constructed so that they satisfy the skew 
orthogonality relations:
\begin{displaymath} \langle R^1_{2j}(x), R^1_{2l+1}(y) \rangle_1 = 
- \langle R^1_{2l+1}(x), R^1_{2j}(y) \rangle_1 = 
r_j \delta_{jl}, \end{displaymath}
\begin{equation} \label{SKEW} \langle R^1_{2j}(x), R^1_{2l}(y) \rangle_1 
= 0, \quad
 \langle R^1_{2j+1}(x), R^1_{2l+1}(y) \rangle_1 = 0. \end{equation}
Defining a set of functions $R^m_k(x), \ \ m=2,3,\cdots,M+1$ as
\begin{equation}
R^m_k(x) = \int_0^{\infty} {\rm d}y R^1_k(y) G^{1 m}(y,x),
\end{equation}
we can immediately find the following skew orthogonality relations 
for $m = 1,2,\cdots,M+1$: 
\begin{displaymath} \langle R^m_{2j}(x), R^m_{2l+1}(y) \rangle_m = 
- \langle R^m_{2l+1}(x), R^m_{2j}(y) \rangle_m = 
r_j \delta_{jl}, \end{displaymath}
\begin{equation} \langle R^m_{2j}(x), R^m_{2l}(y) \rangle_m = 0, \quad
 \langle R^m_{2j+1}(x), R^m_{2l+1}(y) \rangle_m = 0.\end{equation}
At this stage we need to consider the cases $N$ even and odd 
separately.

\subsection{The Case $N$ Even}

Let us now introduce matrices $D^{mn}$, $I^{mn}$, $S^{mn}$, 
$F^{mn}$ and $G^{mn}$ as 
\begin{equation} 
D^{mn}_{jl} =  \sum_{k=0}^{(N/2)-1} \frac{1}{
r_k} [R^m_{2 k}(y^m_j) R^n_{2 k + 1}(y^n_l) - R^m_{2 k + 1}(y^m_j) 
R^n_{2 k}(y^n_l)], 
\end{equation}
\begin{equation} 
I^{mn}_{jl} = - \sum_{k=0}^{(N/2)-1} \frac{1}{
r_k} [ \Phi^m_{2 k}(y^m_j) \Phi^n_{2 k + 1}(y^n_l) 
- \Phi^m_{2 k + 1}(y^m_j) \Phi^n_{2 k}(y^n_l)], 
\end{equation}
\begin{equation}
S^{mn}_{jl} = \sum_{k=0}^{(N/2)-1} 
\frac{1}{r_k} [ \Phi^m_{2 k}(y^m_j) 
R^n_{2 k + 1}(y^n_l) -  \Phi^m_{2 k + 1}(y^m_j) R^n_{2 k}(y^n_l)],  
\end{equation}
where 
\begin{equation}
\Phi^m_k(x) = \int_0^{\infty} F^{mm}(y,x) R^m_k(y) {\rm d}y.
\end{equation}
Further we define
\begin{equation}
F^{mn}_{jl} = F^{mn}(y^m_j,y^n_l),
\label{FMNJL}
\end{equation}
\begin{eqnarray}
G^{mn}_{jl} & = & 0, \ \ m \geq n, \nonumber \\ 
G^{mn}_{jl} & = & G^{mn}(y^m_j,y^n_l), \ \ m < n 
\label{GMNJL}
\end{eqnarray}
and
\begin{equation}
{\tilde I}^{mn}_{jl} = I^{mn}_{jl} + F^{mn}_{jl}, \ \ \ 
{\tilde S}^{mn}_{jl} = S^{mn}_{jl} - G^{mn}_{jl}.
\end{equation}

\subsection{The Case $N$ Odd}

In terms of skew orthogonal functions $R^m_k(x)$, 
another set of functions are defined as
\begin{equation}
{\hat R}^m_k(x) = R^m_k(x) - \frac{\sigma_n}{\sigma_{N-1}} 
R^m_{N-1}(x), \quad n=0,1,\cdots,N-2,
\end{equation}
\begin{equation}
{\hat R}^m_{N-1}(x) = R^m_{N-1}(x),
\end{equation}
where
\begin{equation}
\sigma_k = \int_0^{\infty} {\rm d}x f^M(x) R^M_k(x)
\end{equation}
with 
\begin{equation}
f^m(x) = \int_0^{\infty} {\rm d}y G^{m0}(x,y). 
\end{equation} 
\par
Using the definitions
\begin{equation}
f^m_j = f^m(y^m_j)
\end{equation}
and
\begin{equation}
R^m_j = {\hat R}^m_{N-1}(y^m_j) \frac{1}{\sigma_{N-1}}, 
\end{equation}
we introduce the matrices $D^{mn}$, $I^{mn}$ and $S^{mn}$ 
in the case $N$ odd as 
\begin{equation} 
D^{mn}_{jl} = \sum_{k=0}^{(N-3)/2} \frac{1}{
r_k} [ {\hat R}^m_{2 k}(y^m_j) {\hat R}^n_{2 k + 1}(y^n_l) 
-  {\hat R}^m_{2 k + 1}(y^m_j) {\hat R}^n_{2 k}(y^n_l)],  
\end{equation}
\begin{eqnarray} 
I^{mn}_{jl} & = & - \sum_{k=0}^{(N-3)/2} \frac{1}{
r_k} [ {\hat \Phi}^m_{2 k}(y^m_j) {\hat \Phi}^n_{2 k + 1}(y^n_l) 
-  {\hat \Phi}^m_{2 k + 1}(y^m_j) {\hat \Phi}^n_{2 k}(y^n_l)] 
\nonumber \\ & + & \left[ \frac{1}{\sigma_{N-1}} 
{\hat \Phi}^m_{N-1}(y^m_j) f^n_l - 
\frac{1}{\sigma_{N-1}} {\hat \Phi}^n_{N-1}(y^n_l) f^m_j \right]
\end{eqnarray}
and
\begin{equation}
S^{mn}_{jl} = \sum_{k=0}^{(N-3)/2} \frac{1}{
r_k} [ {\hat \Phi}^m_{2 k}(y^m_j) {\hat R}^n_{2 k + 1}(y^n_l) 
- {\hat \Phi}^m_{2 k + 1}(y^m_j) {\hat R}^n_{2 k}(y^n_l)] + f^m_j R^n_l,
\end{equation}
where
\begin{equation}
{\hat \Phi}^m_k(x) = \int_0^{\infty} F^{mm}(y,x) {\hat R}^m_k(y) {\rm d}y.
\end{equation}
Then the matrices ${\tilde I}^{mn}$ and 
${\tilde S}^{mn}$ in the case $N$ odd are 
given by 
\begin{equation}
{\tilde I}^{mn}_{jl} = I^{mn}_{jl} + F^{mn}_{jl}, \ \ \ 
{\tilde S}^{mn}_{jl} = S^{mn}_{jl} - G^{mn}_{jl},
\end{equation}
where matrices $F^{mn}$ and $G^{mn}$ are defined in eqs. 
($\ref{FMNJL}$) and ($\ref{GMNJL}$). 

\section{Description in terms of the Laguerre Polynomials}
\setcounter{equation}{0}
\renewcommand{\theequation}{3.\arabic{equation}} 
In order to derive the asymptotic behavior of the dynamical 
correlation functions, it is convenient to rewrite the quaternion 
determinant formula presented in \S 2 in terms of the Laguerre 
polynomials. 
Note that 
($t_{M+1} \equiv T$)
\begin{eqnarray}
& & G^{1n}(x,y) \nonumber \\ 
& = & \frac{1}{2 \sqrt{2 \pi t_1}} 
\frac{1}{\sqrt{2 \pi (t_n-t_1) y}} 
{\rm exp}\left\{- \frac{1}{2 t_1} x\right\} \nonumber \\ 
& \times &  
\left[ {\rm exp}\left\{ - \frac{1}{2 (t_n-t_1)} 
(\sqrt{x} - \sqrt{y})^2\right\} - {\rm exp}\left\{ 
- \frac{1}{2 (t_n-t_1)} (\sqrt{x} 
+ \sqrt{y})^2\right\} \right] \nonumber \\  
& = & \frac{1}{\sqrt{2 \pi t_1}} \frac{1}{\sqrt{2 \pi 
(t_n-t_1)}} \frac{1}{2 \sqrt{y}} \nonumber \\ 
& \times & {\rm exp}\left\{- \frac{1}{2} \left( \frac{1}{t_1} + 
\frac{1}{t_n-t_1} \right) x \right\} {\rm exp}\left\{- \frac{1}
{2 (t_n-t_1)} y \right\} \nonumber \\ 
& \times & \left[ 
{\rm exp}\left\{\frac{1}{t_n - t_1} \sqrt{xy}\right\} - 
{\rm exp}\left\{- \frac{1}{t_n - t_1} \sqrt{xy}\right\} \right].
\end{eqnarray} 
Let us define 
\begin{equation}
c_n = \frac{t_n(2 T - t_n)}{T}
\end{equation}
and rescale $x$ and $y$ as
\begin{eqnarray}
& & G^{1n}(c_1 \lambda, c_n \lambda^{\prime}) \nonumber \\ 
& = & \frac{1}{\sqrt{2 \pi t_1}} \frac{1}{\sqrt{2 \pi (t_n-t_1)}}
\frac{1}{2 \sqrt{c_n \lambda^{\prime}}} \exp\left\{ - \frac{1}{1 - z_n} 
(\lambda + \lambda^{\prime}) \right\} \exp\left\{ \frac{t_n}{2 T} \lambda^{\prime} \right\} 
\nonumber \\ 
& \times & \left[ \exp\left\{ \frac{2 \sqrt{z_n \lambda \lambda^{\prime}}}{1 - z_n} 
\right\} -  \exp\left\{ - \frac{2 \sqrt{z_n \lambda \lambda^{\prime}}}{1 - z_n} 
\right\} \right] \nonumber \\ 
& = & \frac{1}{\sqrt{2 \pi t_1}} \frac{1}{\sqrt{2 \pi (t_n-t_1)}}
\sqrt{\frac{\pi}{c_n}} \sqrt{z_n (1 - z_n)} \nonumber \\ 
& \times & 
\lambda^{1/2} {\rm e}^{-\lambda} \exp\left[ - \left( 
1 - \frac{t_n}{2 T} \right) \lambda^{\prime}  \right] 
\sum_{j=0}^{\infty} 
\frac{j! z_n^j}{\Gamma(j + (3/2))}  
L^{(1/2)}_j(\lambda) L^{(1/2)}_j(\lambda^{\prime}). \nonumber \\  
\end{eqnarray} 
Here
\begin{equation}
z_n = \frac{t_1}{2 T - t_1} \frac{2 T - t_n}{t_n}
\end{equation}
and $L_j^{(1/2)}(x)$ are the Laguerre polynomials 
(with a parameter $1/2$). For $m > 1$, we can similarly derive 
\begin{eqnarray}
& & G^{mn}(c_m \lambda, c_n \lambda^{\prime}) = \frac{1}{\sqrt{c_m c_n}} 
\sqrt{\frac{2 \pi (t_m - t_1)}{2 \pi (t_n - t_1)}} 
\sqrt{\frac{z_n (1 - z_n)}{z_m(1 - z_m)}} \nonumber \\ 
& \times & \lambda^{1/2} \exp\left[ - \frac{t_m}{2 T} \lambda \right] 
\exp\left[ - \left(1 - \frac{t_n}{2 T} \right) \lambda^{\prime} \right] 
\nonumber \\  
& \times & \sum_{j=0}^{\infty} \left( \frac{z_n}{z_m} \right)^j 
\frac{j!}{\Gamma(j + (3/2))} L^{(1/2)}_j(\lambda) L^{(1/2)}_j(\lambda^{\prime}). 
\end{eqnarray}
\par 
Let us now define monic orthogonal polynomials
\begin{equation}
C_n(x) = (-1)^n n! L^{(1/2)}_n(x) = x^n + \cdots
\label{ORT}
\end{equation}
satisfying
\begin{equation}
\int_0^{\infty} x^{1/2} {\rm e}^{-x} C_j(x) C_l(x) {\rm d}x = 
h_j \delta_{jl}
\label{ORTH}
\end{equation}
with
\begin{equation}
h_j = (j+(1/2))! j!.
\end{equation}
We then introduce skew orthogonal polynomials
\begin{equation}
{\bar R}^m_k(x) = (\chi_m)^{-k} 
\sum_{r=0}^k  \alpha_{kr} C_r(x) (\chi_m)^r, 
\label{ALPHA}\end{equation}
where
\begin{eqnarray}
\alpha_{2 k \ r} & = & (-1)^r \frac{(2 k)!}{\sqrt{\pi}} 
\frac{(2 k - r - (1/2))!}{(2 k - r)! r!}, \nonumber \\ 
\alpha_{2 k+1 \ r} & = & (-1)^r \frac{(2 k+1)!}{\sqrt{\pi}} 
\left( 2 k - r + \frac{1}{4} \right) 
\frac{(2 k - r - (3/2))!}{(2 k - r + 1)! r!}
\end{eqnarray}
and
\begin{equation}
\chi_m = \frac{z_m}{z_{M+1}}.
\end{equation}
It is known that ${\bar R}^m_k(x)$ satisfy skew orthogonality 
relations\cite{NW,FP} 
\begin{displaymath} \langle \langle {\bar R}^m_{2j}(x), 
{\bar R}^m_{2l+1}(y) \rangle \rangle_m = 
- \langle \langle {\bar R}^m_{2l+1}(x), 
{\bar R}^m_{2j}(y) \rangle \rangle_m = 
{\bar r}^m_j \delta_{jl}, \end{displaymath}
\begin{equation} \langle \langle {\bar R}^m_{2j}(x), 
{\bar R}^m_{2l}(y) \rangle \rangle_m = 0, \quad
 \langle \langle {\bar R}^m_{2j+1}(x), {\bar R}^m_{2l+1}(y) 
\rangle \rangle_m = 0 \end{equation}
with 
\begin{equation}
{\bar r}^m_j = 4 (2 j + 1)! (2 j)! (\chi_m)^{-4 j-1}.
\end{equation} 
Here the antisymmetric inner product is defined as
\begin{eqnarray}
\langle \langle f(x), g(y) \rangle \rangle_m & = &  
\frac{1}{2} \int_0^{\infty} {\rm d}x \int_0^{\infty} {\rm d}y 
x^{1/4} y^{1/4} {\rm e}^{-x/2} {\rm e}^{-y/2} 
{\bar F}^{mm}(y,x) \nonumber \\ 
& \times & [f(y) g(x) - f(x) g(y)],
\end{eqnarray}
where
\begin{eqnarray}
& & {\bar F}^{mn}(x,y) = \int_0^{\infty} {\rm d}z^{\prime} 
\int_0^{z^{\prime}} {\rm d}z \frac{1}{(z z^{\prime})^{1/4}} \nonumber \\ 
& \times & [g^{m \ M+1}(x,z) g^{n \ M+1}(y,z^{\prime}) - 
g^{n \ M+1}(y,z) g^{m \ M+1}(x,z^{\prime})]
\label{FMN}
\end{eqnarray}   
with
\begin{equation}
g^{m n}(x,y) = x^{1/4} y^{1/4} {\rm e}^{-x/2} {\rm e}^{-y/2} 
\sum_{j=0}^{\infty} \frac{C_j(x) C_j(y)}{h_j} \left( 
\frac{z_n}{z_m} \right)^j.
\label{GMN}
\end{equation}
We can readily find ($m>1$)
\begin{eqnarray}
& & G^{1 \ M+1}(c_1 \lambda, c_{M+1} \lambda^{\prime}) = \frac{1}{\sqrt{2 \pi T}} 
\frac{1}{2 T - t_1} \left( \frac{\lambda}{\lambda^{\prime}} \right)^{1/4} 
{\rm e}^{-\lambda/2} g^{1 \ M+1}(\lambda,\lambda^{\prime}), \nonumber \\ 
& & G^{m \ M+1}(c_m \lambda, c_{M+1} \lambda^{\prime}) = \frac{1}{\sqrt{c_m c_{M+1}}} 
\sqrt{\frac{t_m - t_1}{t_{M+1} - t_1}} \sqrt{\frac{z_{M+1} (1 - z_{M+1})}{
z_m ( 1 - z_m )}} \nonumber \\ & \times & 
\left( \frac{\lambda}{\lambda^{\prime}} \right)^{1/4} 
{\rm exp}\left[\left(1 - \frac{t_m}{T}\right)\frac{\lambda}{2}\right] 
g^{m \ M+1}(\lambda,\lambda^{\prime}),
\end{eqnarray}
so that
\begin{eqnarray}
& & F^{11}(c_1 \lambda, c_1 \lambda^{\prime}) = \frac{1}{2 \pi T} 
\left(\frac{t_1}{c_1}\right)^2 (\lambda \lambda^{\prime})^{1/4} {\rm e}^{-\lambda/2} 
{\rm e}^{-\lambda^{\prime}/2} {\bar F}^{11}(\lambda,\lambda^{\prime}), \nonumber \\ 
& & F^{mm}(c_m \lambda,c_m \lambda^{\prime}) =  
\frac{C_{M+1}}{c_m} \frac{t_m - t_1}{t_{M+1}-t_1} \frac{z_{M+1}(1 - z_{M+1})}{
z_m(1 - z_m)} \nonumber \\ 
& \times & (\lambda \lambda^{\prime})^{1/4} 
{\rm exp}\left[\left(1 - \frac{t_m}{T}\right)\frac{\lambda}{2}\right] 
{\rm exp}\left[\left(1 - \frac{t_m}{T}\right)\frac{\lambda^{\prime}}{2}\right] 
{\bar F}^{mm}(\lambda,\lambda^{\prime}).
\end{eqnarray}
It is now straightforward to see that the polynomials
\begin{equation}
R^1_k(x) = (c_1)^k {\bar R}^1_k(x/c_1)
\end{equation}
satisfy the skew orthogonality relation (\ref{SKEW}) with 
\begin{equation}
r_j = \frac{2}{\pi} (2 j + 1)! (2 j)! 
\left(\frac{(t_1)^2}{T} \right)^{4 j + 2},  
\end{equation} 
since
\begin{equation}
\langle \langle {\bar R}^m_j(x), {\bar R}^m_l(y) \rangle \rangle_m 
= 2 \pi \frac{T}{(t_1)^2} (c_1 z_m)^{-j-l} \langle 
R^m_j(x), R^m_l(y) \rangle_m.
\end{equation}
Then, from the definition, we obtain ($m>1$)
\begin{eqnarray}
& & R^m_l(x) = (z_m)^k (c_1)^{k+1} \frac{1}{\sqrt{2 \pi t_1}} \frac{1}{
\sqrt{2 \pi (t_m - t_1)}} \sqrt{\frac{\pi}{c_m}} \sqrt{z_m(1 - z_m)} 
\nonumber \\ 
& \times & \exp\left[-\left(1 - \frac{t_m}{2 T}\right)\frac{x}{c_m} 
\right] {\bar R}^m_k(x/c_m).
\end{eqnarray} 
Putting the above results into the quaternion determinant formula 
yields
\begin{eqnarray}
& & \rho(y^1_1,\cdots,y^1_{n_1};y^2_1,\cdots,y^2_{n_2};\cdots;y^M_1,
\cdots,y^M_{n_M}) \nonumber \\ 
& = & {\rm Tdet}[B^{\mu \nu}(n_{\mu},n_{\nu})] 
\nonumber \\ & = & \prod_{l=1}^M (c_l)^{-n_l} 
{\rm Tdet}[{\bar B}^{\mu \nu}(n_{\mu},n_{\nu})],
\label{BARB}
\end{eqnarray}
where each block ${\bar B}^{\mu \nu}(n_{\mu}, n_{\nu}) $ is obtained by 
removing the $n_{\mu} + 1, n_{\mu} + 2,  \cdots, N$-th rows and 
$n_{\nu} + 1, n_{\nu} + 2, \cdots, N$-th columns from 
${\bar B}^{\mu \nu}$. The quaternion elements ${\bar B}^{\mu \nu}_{jl}$ 
are represented in terms of the scaled variables
\begin{equation}
\lambda^m_j = y^m_j/c_m
\end{equation}
as
\begin{equation}
{\bar B}^{\mu \nu}_{jl} =  \left[ \begin{array}{cc} 
{\bar S}^{\mu \nu} (\lambda^{\mu}_j,\lambda^{\nu}_l) 
& {\bar I}^{\mu \nu}(\lambda^{\mu}_j,\lambda^{\nu}_l) 
 \\ {\bar D}^{\mu \nu}(\lambda^{\mu}_j,\lambda^{\nu}_l) 
& {\bar S}^{\nu \mu}(\lambda^{\nu}_l,\lambda^{\mu}_j) 
\end{array} \right], 
\ \ j,l = 1,2,\cdots,N.
\end{equation}
The definitions of ${\bar S}^{\mu \nu}(x,y)$, 
${\bar I}^{\mu \nu}(x,y)$ and ${\bar D}^{\mu \nu}(x,y)$ 
are (for even $N$)
\begin{eqnarray} 
{\bar D}^{mn}(x,y) & = &   
x^{1/4} {\rm e}^{-x/2} 
y^{1/4} {\rm e}^{-y/2} 
\sum_{k=0}^{(N/2)-1} \frac{1}{
{\bar r}^{M+1}_k} \left[ 
(\chi_m)^{2 k} 
{\bar R}^m_{2 k}(x) 
(\chi_n)^{2 k + 1} 
{\bar R}^n_{2 k + 1}(y) \right. \nonumber \\ 
& - & \left. (\chi_m)^{2 k + 1} 
{\bar R}^m_{2 k + 1}(x) (\chi_n)^{2 k} 
{\bar R}^n_{2 k}(y) \right], 
\label{DMN}
\end{eqnarray}
\begin{eqnarray} 
{\bar I}^{mn}(x,y) & = & - \sum_{k=0}^{(N/2)-1} \frac{1}{
{\bar r}^{M+1}_k} \left[ (\chi_m)^{2 k} 
{\bar \Phi}^m_{2 k}(x) (\chi_n)^{2 k + 1} 
{\bar \Phi}^n_{2 k + 1}(y) \right. \nonumber \\  
& - & \left. (\chi_m)^{2 k+1} 
{\bar \Phi}^m_{2 k + 1}(x) (\chi_n)^{2 k + 1} 
{\bar \Phi}^n_{2 k}(y) \right] + {\bar F}^{mn}(x,y), 
\end{eqnarray}
\begin{eqnarray}
{\bar S}^{mn}(x,y) & = & 
y^{1/4} {\rm e}^{-y/2} 
\sum_{k=0}^{(N/2)-1} 
\frac{1}{{\bar r}^{M+1}_k} \left[ 
(\chi_m)^{2 k} 
{\bar \Phi}^m_{2 k}(x) (\chi_n)^{2 k + 1} 
{\bar R}^n_{2 k + 1}(y) \right. \nonumber \\ 
& - &  \left. (\chi_m)^{2 k + 1} 
{\bar \Phi}^m_{2 k + 1}(x) (\chi_n)^{2 k} 
{\bar R}^n_{2 k}(y) \right] 
- {\bar G}^{mn}(x,y), 
\label{SMN} 
\end{eqnarray}
where 
\begin{equation}
{\bar \Phi}^m_k(x) = \int_0^{\infty} {\bar F}^{mm}(y,x) 
y^{1/4} {\rm e}^{-y/2} {\bar R}^m_k(y) {\rm d}y
\end{equation}
and
\begin{eqnarray}
{\bar G}^{mn}(x,y) & = & 0, \ \ m \geq n, \nonumber \\ 
{\bar G}^{mn}(x,y) & = & g^{mn}(x,y), \ \ m < n. 
\end{eqnarray}

\section{Asymptotic Correlations}
\setcounter{equation}{0}
\renewcommand{\theequation}{4.\arabic{equation}} 
Let us consider the asymptotic limit $N \rightarrow \infty$ of 
the dynamical correlation functions. A new result 
should be searched in the neighborhood of the origin, 
since, in the regions far from the origin, 
the asymptotic behavior of the dynamical 
correlations should not be changed by 
the presence of the wall. In order to see the 
asymptotic correlations around the origin, 
we define scaled temporal 
and spatial variables 
$\upsilon_m$ and $X^m_j$ as
\begin{eqnarray}
\displaystyle 
t_m & = & \left(1 - \frac{\upsilon_m}{2 N} \right) T, \nonumber \\ 
\displaystyle 
\frac{y^m_j}{c_m} & = & \lambda^m_j = \frac{X^m_j}{N}.
\label{SCALING}
\end{eqnarray}
Taking the expansion (\ref{ALPHA}) and putting an 
asymptotic formula
\begin{equation}
\lim_{n \rightarrow \infty} L_n^{(a)}(x) 
\sim {\rm e}^{x/2} \left( \frac{n}{x} \right)^{a/2} 
J_a(2 \sqrt{n x})
\label{BESSEL}
\end{equation}
($J_a(x)$ is the Bessel function) which holds uniformly 
for $x=O(1/n)$, we can readily derive
\begin{equation}
{\bar R}^m_{2k}(x/N) \frac{(\chi_m)^{2 k}}{(2 k)!} \sim 
\frac{N \theta^{3/4}}{\sqrt{\pi} x^{1/4}} 
\int_0^1 {\rm d}\eta \frac{\eta^{1/4}}{(1 - \eta)^{1/2}} 
J_{1/2}(2 \sqrt{\theta \eta x}) {\rm e}^{\theta \eta \upsilon_m},
\label{REVEN}
\end{equation} 
where
\begin{equation}
\theta = \frac{2 k}{N}.
\end{equation}
The skew orthogonal polynomials with odd order are rewritten as
\begin{eqnarray}
& & {\bar R}^m_{2 k + 1}(x) \frac{(\chi_m)^{2 k + 1}}{(2 k + 1)!} 
= - \frac{(2 k + (1/2))!}{\sqrt{\pi} (2 k + 1)!} L^{(1/2)}_0(x) 
- \frac{(2 k - (1/2))!}{\sqrt{\pi} (2 k)!} L^{(1/2)}_1(x) \chi_m  
\nonumber \\ & - &  \frac{1}{\sqrt{\pi}} \sum_{r=2}^{2 k + 1} 
\frac{(2 k - r + (1/2))!}{(2 k - r + 1)!} [L^{(1/2)}_r(x) (\chi_m)^r 
- L^{(1/2)}_{r-2}(x) (\chi_m)^{r-2} ].
\end{eqnarray} 
Here the contribution to the asymptotic form comes from 
the last term and is evaluated as
\begin{equation}
{\bar R}^m_{2 k + 1}(x/N) \frac{(\chi_m)^{2 k + 1}}{(2 k + 1)!} 
\sim - \frac{2}{\sqrt{\pi} (x \theta)^{1/4}} 
\int_0^1 {\rm d}\eta \frac{1}{(1 - \eta)^{1/2}} 
\frac{{\rm d}}{{\rm d} \eta} \left[ \eta^{1/4} 
J_{1/2}(2 \sqrt{\theta \eta x}) {\rm e}^{\theta \eta \upsilon_m} \right].
\label{RODD}
\end{equation}
We substitute the asymptotic forms of ${\bar R}^m_k(x)$ into 
(\ref{DMN}) to derive
\begin{eqnarray}
& & {\bar D}^{mn}(x/N,y/N) \sim  
- \frac{N^{3/2}}{4 \pi} \int_0^1 {\rm d}\theta 
\theta^{1/2} 
\nonumber \\ 
& \times & \left[ 
\int_0^1 {\rm d}\eta 
\frac{\eta^{1/4}
J_{1/2}(2 \sqrt{\theta \eta x}) {\rm e}^{\theta \eta \upsilon_m} 
}{(1 - \eta)^{1/2}} 
\int_0^1 {\rm d}\xi 
\frac{1}{(1 - \xi)^{1/2}} 
\frac{{\rm d}}{{\rm d}\xi} \left\{ \xi^{1/4} 
J_{1/2}(2 \sqrt{\theta \xi y}) {\rm e}^{\theta \xi \upsilon_n} 
\right\} \right. \nonumber \\ 
& - & \left. 
\int_0^1 {\rm d}\eta 
\frac{\eta^{1/4}
J_{1/2}(2 \sqrt{\theta \eta y}) {\rm e}^{\theta \eta \upsilon_n}
}{(1 - \eta)^{1/2}} 
\int_0^1 {\rm d}\xi 
\frac{1}{(1 - \xi)^{1/2}} 
\frac{{\rm d}}{{\rm d}\xi} \left\{ \xi^{1/4} 
J_{1/2}(2 \sqrt{\theta \xi x}) {\rm e}^{\theta \xi \upsilon_m} 
\right\} \right]. \nonumber \\
\label{DASY} 
\end{eqnarray}
\par
We now introduce the inverse matrix $[\beta_{jl}]_{j,l=1,\cdots,k}$ of 
$[\alpha_{jl}]_{j,l=1,\cdots,k}$ as  
\begin{equation}
C_k(x) = (\chi_m)^{-k} \sum_{r=0}^k \beta_{kr} {\bar R}^m_r(x) (\chi_m)^r,
\label{BETA} 
\end{equation}
where
\begin{eqnarray}
\beta_{k \ 2 r} & = & \frac{(-1)^{k+1}}{2 \sqrt{\pi}} 
\frac{(k - 2 r - (3/2))!}{(k - 2 r)!} \frac{k!}{(2 r)!}, 
\nonumber \\ 
\beta_{k \ 2 r + 1} & = & \frac{(-1)^k}{2 \sqrt{\pi}} 
\frac{k!}{(2 r + 1)!}
\sum_{l=0}^{[(k-2r-1)/2]} 
\frac{(k - 2 r - 2 l - (5/2))!}{(k - 2 r - 2 l - 1)!}. 
\end{eqnarray}
In terms of the inverse matrix $\beta_{jl}$, the function 
${\bar \Phi}^m_k(x)$ is written as 
\begin{eqnarray}
{\bar \Phi}^m_{2k+1}(x) (\chi_m)^{2k+1} & = & 
- x^{1/4} {\rm e}^{-x/2} {\bar r}^{M+1}_k
\sum_{\nu = 2 k}^{\infty} \frac{C_{\nu}(x) 
(\chi_m)^{-\nu}}{h_{\nu}} \beta_{\nu \ 2 k}, 
\nonumber \\
{\bar \Phi}^m_{2k}(x) (\chi_m)^{2k} & = & 
x^{1/4} {\rm e}^{-x/2} {\bar r}^{M+1}_k
\sum_{\nu = 2 k + 1}^{\infty} \frac{C_{\nu}(x) 
(\chi_m)^{-\nu}}{h_{\nu}} \beta_{\nu \ 2 k + 1}. \nonumber \\  
\label{PHIBETA}
\end{eqnarray}
Using the above expression and identities 
\begin{eqnarray}
\sum_{l=0}^n \frac{\displaystyle 
\left(2 l - \frac{1}{2} \right)!}{(2 l + 1)!} 
& = & \sqrt{2 \pi} - \sum_{l=n+1}^{\infty} \frac{\displaystyle 
\left(2 l - \frac{1}{2} \right)!}{(2 l + 1)!}, \nonumber \\ 
\sum_{l=0}^n \frac{\displaystyle 
\left(2 l - \frac{3}{2} \right)!}{(2 l)!} 
& = & - \sqrt{2 \pi} - \sum_{l=n+1}^{\infty} \frac{\displaystyle 
\left(2 l - \frac{3}{2} \right)!}{(2 l)!},
\end{eqnarray}
we can rewrite the function ${\bar \Phi}^m_{2 k}(x)$ as
\begin{eqnarray}
& & {\bar \Phi}^m_{2 k}(x) \frac{(\chi_m)^{2 k}}{(2 k)!} 
x^{-1/4} {\rm e}^{x/2} 
\nonumber \\ 
& = & - \frac{2}{\sqrt{\pi}} 
\sum_{r=k+1}^{\infty} \frac{(2 r)!}{(2 r + (1/2))!} 
L^{(1/2)}_{2 r}(x) (\chi_m)^{- 2 r} \sum_{l=r-k}^{\infty} 
\frac{(2 l - (1/2))!}{(2 l + 1)!} \nonumber \\ 
& - & \frac{2}{\sqrt{\pi}} 
\sum_{r=k+1}^{\infty} \frac{(2 r - 1)!}{(2 r - (1/2))!} 
L^{(1/2)}_{2 r - 1}(x) (\chi_m)^{- 2 r + 1} \sum_{l=r-k}^{\infty} 
\frac{(2 l - (3/2))!}{(2 l)!} 
\nonumber \\ 
& + & 2^{3/2} 
\sum_{r=k+1}^{\infty} \left[ \frac{(2 r)!}{(2 r + (1/2))!} 
\frac{L^{(1/2)}_{2 r}(x)}{(\chi_m)^{2 r}} - \frac{(2 r - 1)!}{(2 r - (1/2))!} 
\frac{L^{(1/2)}_{2 r-1}(x)}{(\chi_m)^{2 r - 1}} \right]. \nonumber \\  
\end{eqnarray} 
The first and second terms contribute to the asymptotic limit 
and give
\begin{equation}
{\bar \Phi}^m_{2 k}(x/N) \frac{(\chi_m)^{2 k}}{(2 k)!} 
\sim - \frac{2}{\sqrt{\pi}} (N \theta)^{1/4} \int_1^{\infty} 
\frac{1}{\eta^{1/4} (\eta - 1)^{1/2}} J_{1/2}(2 \sqrt{\theta \eta x}) 
{\rm e}^{- \theta \eta \upsilon_m}.
\label{PHIEVEN}
\end{equation} 
Then we put ${\bar \Phi}^m_{2 k + 1}(x)$ in a form
\begin{eqnarray}
& & {\bar \Phi}^m_{2 k+1}(x) \frac{(\chi_m)^{2 k+1}}{(2 k+1)!} 
\frac{\sqrt{\pi}}{4} x^{-1/4} {\rm e}^{x/2} 
\nonumber \\ 
& = & \sum_{r=k+1}^{\infty} 
\frac{(2 r - 2 k - (5/2))!}{
(2 r - 2 k - 2)!} \frac{(2 r - 2)!}{(2 r - (1/2))!} 
\nonumber \\ & \times & (\chi_m)^{- 2 r + 1}  
\left[ (2 r - 1) L^{(1/2)}_{2 r - 1}(x) 
- \left( 2 r - \frac{1}{2} \right) 
L^{(1/2)}_{2 r - 2}(x) \right] \nonumber \\  
& + & \sum_{r=k+1}^{\infty} 
\frac{(2 r - 2 k - (3/2))!}{
(2 r - 2 k - 1)!} \frac{(2 r - 1)!}{(2 r + (1/2))!} 
\nonumber \\ & \times & (\chi_m)^{- 2 r + 1}  
\left[ (2 r) L^{(1/2)}_{2 r}(x) 
- \left( 2 r + \frac{1}{2} \right) 
L^{(1/2)}_{2 r - 1}(x) \right] 
\nonumber \\  
& + & \sum_{r=k+1}^{\infty} 
\frac{(2 r - 2 k - (7/2))!}{
(2 r - 2 k - 3)!} \frac{(2 r - 2)!}{(2 r - (3/2))!} 
L^{(1/2)}_{2 r - 2}(x) [(\chi_m)^{- 2 r + 2} - (\chi_m)^{- 2 r + 3}]
\nonumber \\  
& + & \sum_{r=k+1}^{\infty} 
\frac{(2 r - 2 k - (5/2))!}{
(2 r - 2 k - 2)!} \frac{(2 r - 2)!}{(2 r - (3/2))!} 
L^{(1/2)}_{2 r - 2}(x) [(\chi_m)^{- 2 r + 1} - (\chi_m)^{- 2 r + 2}] 
\nonumber \\ 
\end{eqnarray}
and substitute (\ref{BESSEL}) to derive 
\begin{equation}
{\bar \Phi}^m_{2 k+1}(x/N) \frac{(\chi_m)^{2 k+1}}{(2 k+1)!} 
\sim \frac{4}{\sqrt{\pi}} \frac{1}{(N \theta)^{3/4}} \int_1^{\infty} 
{\rm d}\eta \frac{1}{(\eta - 1)^{1/2}} \frac{{\rm d}}{{\rm d} \eta} 
\left[ \frac{1}{\eta^{1/4}} J_{1/2}(2 \sqrt{\theta \eta x}) {\rm e}^{-
\theta \eta \upsilon_m} \right].
\label{PHIODD}
\end{equation} 
Putting the expansion (\ref{BETA}) into (\ref{FMN}) and 
using (\ref{PHIBETA}) yields
\begin{eqnarray} 
{\bar F}^{mn}(x,y) & = & \sum_{k=0}^{\infty} \frac{1}{
{\bar r}^{M+1}_k} \left[ (\chi_m)^{2 k} 
{\bar \Phi}^m_{2 k}(x) (\chi_n)^{2 k + 1} 
{\bar \Phi}^n_{2 k + 1}(y) \right. \nonumber \\  
& - & \left. (\chi_m)^{2 k+1} 
{\bar \Phi}^m_{2 k + 1}(x) (\chi_n)^{2 k + 1} 
{\bar \Phi}^n_{2 k}(y) \right], 
\end{eqnarray}
so that
\begin{eqnarray} 
{\bar I}^{mn}(x,y) & = & \sum_{k=N/2}^{\infty} \frac{1}{
{\bar r}^{M+1}_k} \left[ (\chi_m)^{2 k} 
{\bar \Phi}^m_{2 k}(x) (\chi_n)^{2 k + 1} 
{\bar \Phi}^n_{2 k + 1}(y) \right. \nonumber \\  
& - & \left. (\chi_m)^{2 k+1} 
{\bar \Phi}^m_{2 k + 1}(x) (\chi_n)^{2 k + 1} 
{\bar \Phi}^n_{2 k}(y) \right].
\end{eqnarray}
Asymptotic forms (\ref{PHIEVEN}) and (\ref{PHIODD}) are substituted 
into the above expression and yield
\begin{eqnarray}
& & {\bar I}^{mn}(x/N,y/N) \sim - \frac{N^{1/2}}{\pi} \int_1^{\infty} 
{\rm d}\theta \frac{1}{\theta^{1/2}} \nonumber \\ 
& \times & \left[ \int_1^{\infty} {\rm d}\eta \frac{
J_{1/2}(2 \sqrt{\theta \eta x}) {\rm e}^{-\theta \eta \upsilon_m} 
}{\eta^{1/4} 
(\eta - 1)^{1/2}} 
\int_1^{\infty} {\rm d}\xi \frac{1}{(\xi - 1)^{1/2}} 
\frac{{\rm d}}{{\rm d}\xi} 
\left\{ \frac{1}{\xi^{1/4}} J_{1/2}(2 \sqrt{\theta \xi y}) 
{\rm e}^{- \theta \xi \upsilon_n} \right\} \right. \nonumber \\  
& - & \left. \int_1^{\infty} {\rm d}\eta \frac{
J_{1/2}(2 \sqrt{\theta \eta y}) {\rm e}^{-\theta \eta \upsilon_n} 
}{\eta^{1/4} 
(\eta - 1)^{1/2}} 
\int_1^{\infty} {\rm d}\xi \frac{1}{(\xi - 1)^{1/2}} 
\frac{{\rm d}}{{\rm d}\xi} 
\left\{ \frac{1}{\xi^{1/4}} J_{1/2}(2 \sqrt{\theta \xi x}) 
{\rm e}^{- \theta \xi \upsilon_m} \right\} \right]. \nonumber \\  
\label{IASY} 
\end{eqnarray}
\par
Moreover we put (\ref{REVEN}), (\ref{RODD}), (\ref{PHIEVEN}) 
and (\ref{PHIODD}) into (\ref{SMN}) and find 
\begin{eqnarray}
& & {\bar S}^{mn}(x/N,y/N) + {\bar G}^{mn}(x/N,y/N) \sim 
\frac{N}{2 \pi} \int_0^1 {\rm d}\theta \nonumber \\ 
& \times & \left[ \int_1^{\infty} {\rm d}\eta \frac{
J_{1/2}(2 \sqrt{\theta \eta x}) {\rm e}^{-\theta \eta \upsilon_m} 
}{\eta^{1/4} 
(\eta-1)^{1/2}} 
\int_0^1 {\rm d}\xi \frac{1}{(1 - \xi)^{1/2}} \frac{{\rm d}}{{\rm d} \xi} 
\left\{ \xi^{1/4} J_{1/2}(2 \sqrt{\theta \xi y}) {\rm e}^{\theta \xi 
\upsilon_n} \right\} 
\right.  \nonumber \\ 
& - & \left. \int_1^{\infty} {\rm d}\eta \frac{1}{(\eta - 1)^{1/2}} 
\frac{{\rm d}}{{\rm d} \eta} \left\{ \frac{1}{\eta^{1/4}} J_{1/2}(
2 \sqrt{\theta \eta x}) {\rm e}^{-\theta \eta \upsilon_m} \right\} \int_0^1 {\rm d}\xi 
\frac{\xi^{1/4} J_{1/2}(2 \sqrt{\theta \xi y}) 
{\rm e}^{\theta \xi \upsilon_n}}{(1 - \xi)^{1/2}} 
\right]. \nonumber \\  
\label{SASY}
\end{eqnarray}
The function ${\bar G}^{mn}(x,y)$ is identical to $g^{mn}(x,y)$ when $m<n$. 
The asymptotic limit of $g^{mn}(x,y)$ is derived from 
(\ref{GMN}) and (\ref{BESSEL}) as
\begin{equation}
g^{mn}(x/N,y/N) \sim N \int_0^{\infty} {\rm d}s 
J_{1/2}(2 \sqrt{s x}) J_{1/2}(2 \sqrt{s y}) {\rm e}^{- (\upsilon_m 
- \upsilon_n) s}.
\label{GASY}
\end{equation}
Substituting (\ref{DASY}),(\ref{IASY}),(\ref{SASY}) and (\ref{GASY}) into 
(\ref{BARB}), we can see how the dynamical correlation functions 
asymptotically depend on the scaled variables $\upsilon_m$ and $X^m_j$.

\section{Bogoliubov-deGennes Matrix Model}
\setcounter{equation}{0}
\renewcommand{\theequation}{5.\arabic{equation}} 
In this last section we show an equivalence relation between 
the vicious walk model with a wall and the Bogoliubov-deGennes 
matrix model describing the symmetry crossover $CI \rightarrow C$. 
The Bogoliubov-deGennes matrix model was proposed by Altland and 
Zirnbauer as a model of normalconducting-superconducting hybrid 
structures in mesoscopic physics\cite{AZ1,AZ2}. It is a part 
of a classification scheme of random matrix ensembles in 
terms of the Lie algebra\cite{MC,ZIRN}. In the class $C$, 
the spin is conserved while the time reversal symmetry is broken. 
Then the (reduced) system Hamiltonian has a structure
\begin{equation}
{\cal H}_C = \left( \begin{array}{cc} a & b \\ b^{\dagger} 
& - a^{\rm T} \end{array} \right)
\label{HC} 
\end{equation}
with an $N \times N$ hermitian $a$ and an $N \times N$ complex 
symmetric $b$.  On the other hand, in the class $CI$, the system is 
symmetric with respect to both spin rotations and 
time reversal. Then the Hamiltonian matrix structure is
\begin{equation}
{\cal H}_{CI} = \left( \begin{array}{cc} a & b \\ 
b & - a^{\rm T} \end{array} \right), 
\label{HC1} 
\end{equation}
where $a$ and $b$ are both $N \times N$ real symmetric matrices. 
\par
Dyson proposed Brownian motion models for parametric random 
matrices\cite{DYB}. In his prescription, the time evolution 
of a combination 
\begin{equation}
{\cal H} = {\rm e}^{-\tau} ({\cal H}_{CI} + \sqrt{{\rm e}^{2 \tau} - 1} {\cal H}_C )
\end{equation}
describes the crossover $CI \rightarrow C$. The matrix ${\cal H}$ is 
identical to ${\cal H}_{CI}$ at $\tau = 0$ while ${\cal H}$ 
approaches ${\cal H}_C$ as $\tau$ goes to infinity.  
Let us redefine $a$ and $b$ as 
\begin{equation}
{\cal H} = \left( \begin{array}{cc} a & b \\ b^{\dagger} 
& - a^{\rm T} \end{array} \right) 
\end{equation}
so as to reproduce (\ref{HC}) in the limit $\tau \rightarrow \infty$.  
We can diagonalize ${\cal H}$ as
\begin{equation}
{\cal H} = U^{\dagger} \left( \begin{array}{cc} \omega & 0 \\ 
0 & -\omega \end{array} \right) U,
\end{equation}
where $U$ is a $2 N \times 2 N$ unitary matrix and $\omega$ is 
a real diagonal matrix 
\begin{equation}
\omega = \left( \begin{array}{cccc} 
\omega_1 & 0 & \cdots & 0 \\  
0  & \omega_2 & \ddots & 0 \\  
\vdots  & \vdots & \ddots & \vdots \\  
0  & \cdots & \cdots & \omega_N \end{array} 
\right).
\end{equation}
Assuming that ${\cal H}_C$ is distributed according to the Gaussian 
distribution, we obtain the probability distribution function for 
${\cal H}$ as   
\begin{equation}
P({\cal H};\tau) {\rm d}{\cal H}= A_{\tau} {\rm exp}\left[-\frac{{\rm Tr} 
({\cal H} - {\rm e}^{- \tau} {\cal H}_{CI})^2}{2(1 - {\rm e}^{- 
2 \tau})}\right] {\rm d}{\cal H},
\end{equation}
where
\begin{equation} 
{\rm d}{\cal H} = \prod_{j=1}^N {\rm d}a_{jj} {\rm d}b^R_{jj} {\rm d}b^I_{jj} 
\prod_{j<l}^N {\rm d}a^R_{jl} {\rm d}a^I_{jl} {\rm d}b^R_{jl} {\rm d}b^I_{jl}.
\end{equation} 
Here the normalization constant $A_{\tau}$ is evaluated as
\begin{equation}
A_{\tau} = 2^{N(N-1)} \pi^{- N^2-(N/2)} ( 1 - {\rm e}^{- 2 \tau}) 
^{-N^2-(N/2)}.
\end{equation}
\par
It is easy to calculate the differentiations of $P({\cal H};\tau)$ and find 
the Fokker-Planck equation
\begin{eqnarray}
& & \frac{\partial P}{\partial \tau} = \Delta P 
+ \sum_{j=1}^N \left[ 
\frac{\partial}{\partial a_{jj}} (a_{jj} P) +  
\frac{\partial}{\partial b^R_{jj}} (b^R_{jj} P) +  
\frac{\partial}{\partial b^I_{jj}} (b^I_{jj} P) \right] \nonumber \\  
& + & \sum_{j<l}^N \left[ 
\frac{\partial}{\partial a^R_{jl}} (a^R_{jl} P) +  
\frac{\partial}{\partial a^I_{jl}} (a^I_{jl} P) + 
\frac{\partial}{\partial b^R_{jl}} (b^R_{jl} P) + 
\frac{\partial}{\partial b^I_{jl}} (b^I_{jl} P) \right],
\end{eqnarray} 
where $\Delta$ is the Laplace-Beltrami operator
\begin{equation}
\Delta = \frac{1}{2} 
\sum_{j=1}^N \left[ 
\frac{\partial^2}{\partial (a_{jj})^2} +  
\frac{\partial^2}{\partial (b^R_{jj})^2} +  
\frac{\partial^2}{\partial (b^I_{jj})^2} \right] 
 + \frac{1}{4} \sum_{j<l}^N \left[ 
\frac{\partial^2}{\partial (a^R_{jl})^2} +  
\frac{\partial^2}{\partial (a^I_{jl})^2} +  
\frac{\partial^2}{\partial (b^R_{jl})^2} +  
\frac{\partial^2}{\partial (b^I_{jl})^2} \right].
\end{equation}
The Laplace-Beltrami operator on a Riemannian 
manifold defined by the line element
\begin{equation}
{\rm d}s^2 = \sum_{\mu \nu} g_{\mu \nu} {\rm d}x^{\mu} {\rm d}x^{\nu}
\label{LELE}
\end{equation}
is given by 
\begin{equation}
\Delta = \sum_{\mu \nu} \frac{1}{\sqrt{|\det g|}} \frac{\partial}{\partial x^{\mu}} 
(g^{-1})_{\mu \nu} \sqrt{|\det g|} \frac{\partial}{\partial x^{\nu}}.
\label{LBO}
\end{equation}
In our case the corresponding line element is 
\begin{eqnarray}
{\rm d}s^2 & = & 2 \sum_{j=1}^N \left[ 
({\rm d}a_{jj})^2 + 
({\rm d}b^R_{jj})^2 + 
({\rm d}b^I_{jj})^2 \right]
+ 4 \sum_{j<l}^N  \left[
({\rm d}a^R_{jl})^2 + 
({\rm d}a^I_{jl})^2 + 
({\rm d}b^R_{jl})^2 + 
({\rm d}b^I_{jl})^2 \right] \nonumber \\ 
& = & 2 \sum_{j=1}^N \sum_{l=1}^N \left[ {\rm d}a_{jl} {\rm d}a_{jl}^* + 
{\rm d}b_{jl} {\rm d}b_{jl}^* \right].
\end{eqnarray}
Let us introduce $N \times N$ matrices $u_1,u_2,u_3$ and $u_4$ as 
\begin{equation}
U = \left( \begin{array}{cc} u_1 & u_2 \\ u_3 &  u_4 \end{array} \right)
\end{equation}
and rewrite ${\rm d}a_{jl}$ and ${\rm d}b_{jl}$ as 
\begin{eqnarray} 
{\rm d}a_{jl} & = & \sum_{k=1}^N  
\left[ (u_1)_{kj}^* (u_1)_{kl} - (u_3)_{kj}^* (u_3)_{kl} \right] 
{\rm d}\omega_k + {\rm d}A_{jl}, \nonumber \\  
{\rm d}b_{jl} & = & \sum_{k=1}^N  
\left[ (u_1)_{kj}^* (u_2)_{kl} - (u_3)_{kj}^* (u_4)_{kl} \right] 
{\rm d}\omega_k + {\rm d}B_{jl},
\end{eqnarray} 
where
\begin{eqnarray}
{\rm d}A_{jl} & = & \sum_{k=1}^N \left[ ({\rm d}u_1)_{kj}^* (u_1)_{kl} 
+ (u_1)_{kj}^* ({\rm d}u_1)_{kl} - ({\rm d}u_3)_{kj}^* (u_3)_{kl} - 
(u_3)_{kj}^* ({\rm d}u_3)_{kl} \right] \omega_k, \nonumber \\  
{\rm d}B_{jl} & = & \sum_{k=1}^N \left[ ({\rm d}u_1)_{kj}^* (u_2)_{kl} 
+ (u_1)_{kj}^* ({\rm d}u_2)_{kl} - ({\rm d}u_3)_{kj}^* (u_4)_{kl} - 
(u_3)_{kj}^* ({\rm d}u_4)_{kl} \right] \omega_k. \nonumber 
\end{eqnarray}
Using the unitarity of the matrix $U$ ($U^{\dagger} U = U U^{\dagger} = I$), 
we can readily see that
\begin{equation}
{\rm d}s^2 = 2 \sum_{k=1}^N ({\rm d}\omega_k)^2 + 2 \sum_{j=1}^N 
\sum_{l=1}^N \left[ {\rm d}A_{jl} {\rm d}A_{jl}^*  
 + {\rm d}B_{jl} {\rm d}B_{jl}^* \right],
\end{equation}
which, by means of (\ref{LELE}) and (\ref{LBO}), yields
\begin{equation}
\Delta = \frac{1}{2J} \sum_{j=1}^N \frac{\partial}{\partial \omega_j} 
\left( J \frac{\partial}{\partial \omega_j} \right) + \Delta_U.
\end{equation} 
Here an operator $\Delta_U$ involves derivatives with respect to 
the variables associated with $U$. The Jacobian $J = \sqrt{|\det g|}$ 
was evaluated in Ref.\cite{AZ1,AZ2} as
\begin{equation}
J = \prod_{j<l}^N | \omega_j^2 - \omega_l^2 |^2 \prod_{j=1}^N | 
\omega_j |^2.
\end{equation}
Assuming that $P({\cal H};\tau)$ depends only on the radial 
variables $\omega_k$ and a time variable 
$\tau$, we can rewrite the Fokker-Planck equation as
\begin{equation}
\frac{\partial P}{\partial \tau} = \frac{1}{2J} 
\sum_{j=1}^N \frac{\partial}{\partial \omega_j} 
\left( J \frac{\partial P}{\partial \omega_j} \right) + 
\sum_{j=1}^N \omega_j \frac{\partial P}{\partial \omega_j} 
+ (2 N^2 + N) P.
\label{FPOMEGA}
\end{equation}
Substituting $P = p/J$ gives
\begin{equation}
\frac{\partial p}{\partial \tau} = {\cal L}p, \ \ \ 
{\cal L} = \sum_{j=1}^N \frac{\partial}{\partial \omega_j} 
\left( \frac{\partial W}{\partial \omega_j} + \frac{1}{2} \frac{\partial}{
\partial \omega_j} \right) = \frac{1}{2} \sum_{j=1}^N \frac{\partial}{\partial \omega_j} 
{\rm e}^{-2 W} \frac{\partial}{\partial \omega_j} {\rm e}^{2 W},
\end{equation}
where
\begin{equation}
W = \frac{1}{2} \sum_{j=1}^N \omega_j^2 - \frac{1}{2} \log J.
\end{equation}
\par
In order to transform the Fokker-Planck operator into a form 
in which all the variables are separated, let us consider  
\begin{equation}
- \frac{1}{2} (H - E_0) = {\rm e}^W {\cal L} {\rm e}^{-W} 
\end{equation}
($E_0$ is a constant) and find
\begin{equation}
H = - \sum_{j=1}^N \frac{\partial^2}{\partial \omega_j^2} 
+ \sum_{j=1}^N \omega_j^2.
\end{equation}
For the imaginary time Schr\"odinger equation
\begin{equation}
\frac{\partial \psi}{\partial \tau} = - \frac{1}{2} ( H - E_0 ) \psi,
\label{IMSCH}
\end{equation}
we call $\psi = G^{(H)}(\omega^{(0)}_1,\cdots,\omega^{(0)}_N;\omega_1,\cdots,
\omega_N;\tau)$ the Green function solution if it satisfies the 
initial condition
\begin{equation}
G^{(H)}(\omega^{(0)}_1,\cdots,\omega^{(0)}_N;\omega_1,\cdots,
\omega_N;0) = \prod_{j=1}^N \delta(\omega_j - \omega^{(0)}_j).
\end{equation}
Since the imaginary time Schr\"odinger equation (\ref{IMSCH}) describes the 
dynamics of free fermions, the Green function solution is given by  
\begin{equation}
G^{(H)}(\omega^{(0)}_1,\cdots,\omega^{(0)}_N;\omega_1,\cdots,
\omega_N;\tau) = \det[g^{(H)}(\omega^{(0)}_j,\omega_l;
\tau)]_{j,l=1,\cdots,N},
\end{equation}
where $g^{(H)}(\omega^{(0)},\omega; \tau)$ is the Green function 
solution of (\ref{IMSCH}) with $N=1$. 
\par
Let us denote the Green function solution 
of the Fokker-Planck equation (\ref{FPOMEGA}) 
as $G^{(FP)}(\omega^{(0)}_1,\cdots,
\omega^{(0)}_N;\omega_1,\cdots, \omega_N;\tau)$. Then we can readily 
see that
\begin{eqnarray}
& & G^{(FP)}(\omega^{(0)}_1,\cdots,
\omega^{(0)}_N;\omega_1,\cdots, \omega_N;\tau) 
= {\rm e}^{\tau E_0/2} \frac{{\rm e}^{-W(\omega_1,\cdots,\omega_N)}}{
{\rm e}^{-W(\omega^{(0)}_1,\cdots,\omega^{(0)}_N)}}   
G^{(H)}(\omega^{(0)}_1,\cdots,
\omega^{(0)}_N;\omega_1,\cdots, \omega_N;\tau) \nonumber \\ 
& = & {\rm e}^{\tau E_0/2} 
\prod_{j=1}^N (4 \omega_j \omega^{(0)}_j) 
\frac{\displaystyle  
\prod_{j=1}^N |\omega_j|^{1/2} {\rm e}^{-(\omega_j)^2/2} 
\prod_{j>l}^N ((\omega_j)^2 - (\omega_l)^2) }{\displaystyle
\prod_{j=1}^N |\omega^{(0)}_j|^{1/2} {\rm e}^{-(\omega^{(0)}_j)^2/2}
\prod_{j>l}^N ((\omega^{(0)}_j)^2 - (\omega^{(0)}_l)^2)} 
\det[g((\omega^{(0)}_j)^2,(\omega_l)^2;
\tau)]_{j,l=1,\cdots,N}, \nonumber \\ 
\end{eqnarray}
where
\begin{equation}
g(x,y; \tau) = x^{1/4} y^{1/4} {\rm e}^{-x/2} {\rm e}^{-y/2} 
\sum_{j=0}^{\infty} \frac{1}{h_j} C_j(x) C_j(y) {\rm e}^{- 2 j \tau}. 
\end{equation}
Here $C_j(x)$ and $h_j$ are defined in (\ref{ORT}) and 
(\ref{ORTH}), respectively. Note that the above result 
gives the Harish-Chandra integral\cite{HC} in the case 
of the Bogoliubov-deGennes symmetry. We introduce new variables 
$\epsilon_j = \omega_j^2$ and the corresponding 
Green functions as
\begin{equation}
G(\epsilon^{(0)}_1,\cdots,
\epsilon^{(0)}_N;\epsilon_1,\cdots, \epsilon_N;\tau) 
\prod_{j=1}^N 
{\rm d}\epsilon^{(0)}_j {\rm d}\epsilon_j =  
G^{(FP)}(\omega^{(0)}_1,\cdots,
\omega^{(0)}_N;\omega_1,\cdots, \omega_N;\tau) 
\prod_{j=1}^N 
{\rm d}\omega^{(0)}_j {\rm d}\omega_j  
\end{equation}
so that 
\begin{equation}
G(\epsilon^{(0)}_1,\cdots,
\epsilon^{(0)}_N;\epsilon_1,\cdots, \epsilon_N;\tau) 
= {\rm e}^{\tau E_0/2} \frac{\displaystyle  
\prod_{j=1}^N \epsilon_j^{1/4} {\rm e}^{-\epsilon_j/2} 
\prod_{j>l}^N (\epsilon_j - \epsilon_l) }{\displaystyle
\prod_{j=1}^N (\epsilon^{(0)}_j)^{1/4} {\rm e}^{-\epsilon^{(0)}_j/2}
\prod_{j>l}^N (\epsilon^{(0)}_j - \epsilon^{(0)}_l)} 
\det[g(\epsilon^{(0)}_j,\epsilon_l;
\tau)]_{j,l=1,\cdots,N}.  
\end{equation}
\par
The probability distribution functions for the eigenparameters 
$\epsilon^n_j$ at times $\tau_n$ can be evaluated from the 
Green function as
\begin{eqnarray}
& & p(\epsilon^1_1,\cdots,\epsilon^1_N;\tau_1; 
\epsilon^2_1,\cdots,\epsilon^2_N; \tau_2; 
\cdots; \epsilon^M_1,\cdots,\epsilon^M_N; \tau_M) \nonumber 
\\ & = & \frac{1}{N!} \int_0^{\infty} {\rm d}\epsilon^0_1 \cdots 
\int_0^{\infty} {\rm d}\epsilon^0_N p_0(\epsilon^0_1,\cdots,
\epsilon^0_N) \prod_{l=1}^M G(\epsilon^{l-1}_
1,\cdots,\epsilon^{l-1}_N;\epsilon^l_1,\cdots,
\epsilon^l_N;\tau_l - \tau_{l-1}), \nonumber \\ 
\end{eqnarray}
where $p_0(\epsilon_1,\cdots,\epsilon_N)$ 
is the initial probability distribution 
function at $\tau_0 = 0$. The corresponding multilevel dynamical 
correlation functions are given by 
\begin{eqnarray}
& & \rho_{BdG}(\epsilon^1_1,\cdots,\epsilon^1_{m_1};
\epsilon^2_1,\cdots,\epsilon^2_{m_2};
\cdots;\epsilon^M_1,\cdots,\epsilon^M_{m_M}) \nonumber \\ & = & \frac{1}
{C_N} \frac{(N!)^M}{\prod_{l=1}^M (N-m_l)! } \int_0^{\infty} 
{\rm d}\epsilon^1_{m_1+1} \cdots \int_0^{\infty} 
{\rm d}\epsilon^1_N \cdots \int_0^{\infty} {\rm d}\epsilon^M_{m_M+1} \cdots 
\int_0^{\infty} {\rm d}\epsilon^M_N \nonumber 
\\ & \times & p(\epsilon^1_1,\cdots,\epsilon^1_N;\tau_1; 
\epsilon^2_1,\cdots,\epsilon^2_N; \tau_2; 
\cdots; \epsilon^M_1,\cdots,\epsilon^M_N; \tau_M).
\label{RHOBDG} 
\end{eqnarray}
Here we define the normalization constant $C_N$ as 
\begin{eqnarray}
C_N & = & \int_0^{\infty} {\rm d}\epsilon^1_1 \cdots \int_0^{\infty} 
{\rm d} \epsilon^1_N  \cdots \int_0^{\infty} 
{\rm d}\epsilon^M_1 \cdots \int_0^{\infty} {\rm d}\epsilon^M_N 
\nonumber \\ & \times & p(\epsilon^1_1,\cdots,\epsilon^1_N; 
\tau_1; \epsilon^2_1,\cdots,\epsilon^2_N; 
\tau_2; \cdots; \epsilon^M_1,\cdots,\epsilon^M_N; \tau_M).
\end{eqnarray}
The initial eigenparameter distribution for the Gaussian 
random matrices with the symmetry (\ref{HC1}) 
can be written as\cite{AZ2}  
\begin{equation}
p_0(\omega_1,\cdots,\omega_N) 
{\rm d}\omega_1 \cdots {\rm d}\omega_N   
 \propto \prod_j^N {\rm e}^{- \alpha \omega_j^2/2} 
\prod_{j<l}^N | \omega_j^2 - \omega_l^2 | \prod_{j=1}^N | 
\omega_j | {\rm d}\omega_1 \cdots {\rm d}\omega_N
\end{equation}
or, equivalently,    
\begin{equation}
p_0(\epsilon_1,\cdots,\epsilon_N) 
{\rm d}\epsilon_1 \cdots {\rm d}\epsilon_N   
 \propto \prod_j^N {\rm e}^{- \alpha \epsilon_j/2} 
\prod_{j<l}^N | \epsilon_j - \epsilon_l | 
{\rm d}\epsilon_1 \cdots {\rm d}\epsilon_N.
\label{PZERO}   
\end{equation}
A parameter $\alpha$ determines the variance of the 
Gaussian distribution for the matrix elements and we 
set $\alpha = 1$. 
\par
We now see that multilevel dynamical correlation 
functions defined in (\ref{RHOBDG}) with the initial 
condition (\ref{PZERO}) have the same forms as the 
dynamical correlation functions for vicious walkers 
(\ref{RHOINT}). Therefore we can similarly rewrite them 
in quaternion determinant forms\cite{NFQ}
\begin{eqnarray}
& & \rho_{BdG}(\epsilon^1_1,\cdots,\epsilon^1_{m_1};
\epsilon^2_1,\cdots,\epsilon^2_{m_2};
\cdots;\epsilon^M_1,\cdots,\epsilon^M_{m_M}) 
= {\rm Tdet}[{\bar B}^{\mu \nu}(n_{\mu},n_{\nu})], 
\nonumber \\ & & \mu,\nu = 1,2,\cdots,M. 
\end{eqnarray}   
Here the quaternion determinant is identical to 
that in (\ref{BARB}) if we adopt a correspondence
\begin{equation}
\epsilon^l_j=\lambda^{M-l+1}_j, \ \ m_l=n_{M-l+1}, \ \ 
{\rm e}^{2 \tau_l} = \chi_{M-l+1}.   
\end{equation}
Therefore all the dynamical correlation functions 
are shared by the matrix model and the vicious 
walk model. We have thus shown the equivalence 
of the vicious walk model with a wall in the diffusion 
scaling limit and the parametric Bogoliubov-deGennes 
matrix model. This equivalence holds for finite 
$N$ and also in the asymptotic limits $N \rightarrow \infty$. 
Although we have here presented the equivalence only 
for even $N$, we can similarly and straightforwardly prove 
it for odd $N$. 
\par
The eigenparameter distributions of the Bogoliubov-deGennes 
matrix model in the limits $\tau \rightarrow \infty$ and $\tau = 0$ 
are known to be equivalent to the eigenvalue distributions of 
the Laguerre unitary and orthogonal ensembles\cite{NS1,NS2} of 
random matrices, respectively. Let us consider the 
limit $\tau_m \rightarrow \infty$ with the time differences 
$\tau_m-\tau_n$ fixed. We can readily see that in this 
limit the quaternion determinant is reduced to an 
ordinary determinant. The resulting determinant 
expressions describe temporally homogeneous 
dynamical correlations within the $C$ universality class. 
It follows from the rescaling (\ref{SCALING}) that the $C$ 
universality class survives until time $t$ very close 
to $T$: only when $T-t \sim O(N^{-1})$, the transition 
to $CI$ class occurs. Therefore we can conclude that the 
transition from $C$ to $CI$ class is discontinuous in the 
limit $N \rightarrow \infty$. The asymptotic correlation 
functions describing the $CI$ universality class is obtained by 
putting $\upsilon_m = \upsilon_n = 0$ (equivalent to 
$\tau_{M-m+1} = \tau_{M-n+1} = 0$) in (\ref{DASY}), (\ref{IASY}) 
and (\ref{SASY}). Using the Bessel function identities, 
we can easily confirm that they are identical to Nagao and 
Slevin's result\cite{NS2} for the Laguerre orthogonal ensemble.          

\section{Conclusion}
\setcounter{equation}{0}
\renewcommand{\theequation}{6.\arabic{equation}} 
In this paper we have analyzed the vicious walk model 
with a wall in the diffusion scaling limit. It was shown 
that all the dynamical correlation functions are 
written in the forms of quaternion determinants. 
Using the quaternion determinant formulas we were able 
to derive the asymptotic formulas for the correlation 
functions. Finally we showed that the vicious walk 
model in the diffusion scaling limit was equivalent to 
the parametric Bogoliubov-deGennes matrix model. As the 
equivalence to the matrix model is so far established only
in the diffusion scaling limit, it is interesting to 
consider how the matrix model should be generalized 
corresponding to the discrete vicious walk.     

\section*{Acknowledgement}
The author is grateful to Prof. M. Katori, Prof. H. Tanemura and 
Dr. P.J. Forrester for valuable discussions.

\end{document}